\documentclass[journal]{IEEEtran}

\usepackage[T1]{fontenc}% optional T1 font encoding

\usepackage{cite}

\usepackage{multirow}
\usepackage{graphicx}
\usepackage{amssymb}
\usepackage{amsmath}
\usepackage{booktabs}
\usepackage{array}
\newcolumntype{L}[1]{>{\raggedright\let\newline\\\arraybackslash\hspace{0pt}}m{#1}}
\newcolumntype{C}[1]{>{\centering\let\newline\\\arraybackslash\hspace{0pt}}m{#1}}
\newcolumntype{R}[1]{>{\raggedleft\let\newline\\\arraybackslash\hspace{0pt}}m{#1}}

\DeclareMathOperator*{\argmax}{argmax}

\usepackage{lipsum}
\usepackage{changepage}
\usepackage{color,soul}
\usepackage{enumitem}   

\usepackage{pifont}

\usepackage{tikz}

\usepackage{lipsum}

\let\OLDthebibliography\thebibliography
\renewcommand\thebibliography[1]{
  \OLDthebibliography{#1}
  \setlength{\parskip}{0pt}
  \setlength{\itemsep}{0pt plus 0.3ex}
}

\usepackage{array}

\makeatletter
\newcommand{\thickhline}{%
    \noalign {\ifnum 0=`}\fi \hrule height 1pt
    \futurelet \reserved@a \@xhline
}
\newcolumntype{"}{@{\hskip\tabcolsep\vrule width 1pt\hskip\tabcolsep}}
\makeatother

\usepackage[ruled,vlined]{algorithm2e}

\usepackage{framed,multirow}

\usepackage{latexsym}

\usepackage{url}
\usepackage{xcolor}

\usepackage{hyperref}

\usepackage{bm}

\begin{document}

\title{Learning to segment fetal brain tissue from noisy annotations}

\author{Davood Karimi, Caitlin K. Rollins, Clemente Velasco-Annis, Abdelhakim Ouaalam, and Ali Gholipour, \IEEEmembership{Senior Member, IEEE}
\thanks{This project was supported in part by the National Institute of Biomedical Imaging and Bioengineering, the National Institute of Neurological Disorders and Stroke, and the Eunice Kennedy Shriver National Institute of Child Health and Human Development of the National Institutes of Health (NIH) under award numbers R01EB031849, R01EB032366, R01HD109395, R01NS106030, K23NS101120, and R01NS121334; in part by the Office of the Director of the NIH under award number S10OD0250111; in part by the National Science Foundation (NSF) under award 2123061; in part by a research award from the Additional Ventures; in part by NVIDIA corporation through the Applied Research Accelerator Program and the Academic Hardware Grant Program; and in part by a Technological Innovations in Neuroscience Award from the McKnight Foundation. The content of this paper is solely the responsibility of the authors and does not necessarily represent the official views of the NIH, NSF, Additional Ventures, or the McKnight Foundation.}
\thanks{D. Karimi, C. Velasco-Annis, A. Ouaalam, and A. Gholipour are with the Department of Radiology, Boston Children's Hospital, and Harvard Medical School, Boston, Massachusetts, USA.}
\thanks{C.K. Rollins is with the Department of Neurology at Boston Children's Hospital, and Harvard Medical School, Boston, Massachusetts, USA.}
\thanks{Corresponding author: D. Karimi, davood.karimi@childrens.harvard.edu.}
\thanks{Code and trained models for this project can be found at: https://github.com/bchimagine/fetal\_tissue\_segmentation.}
}

\maketitle

\begin{abstract}
Automatic fetal brain tissue segmentation can enhance the quantitative assessment of brain development at this critical stage. Deep learning methods represent the state of the art in medical image segmentation and have also achieved impressive results in brain segmentation. However, effective training of a deep learning model to perform this task requires a large number of training images to represent the rapid development of the transient fetal brain structures. On the other hand, manual multi-label segmentation of a large number of 3D images is prohibitive. To address this challenge, we segmented 272 training images, covering 19-39 gestational weeks, using an automatic multi-atlas segmentation strategy based on deformable registration and probabilistic atlas fusion, and manually corrected large errors in those segmentations. Since this process generated a large training dataset with noisy segmentations, we developed a novel label smoothing procedure and a loss function to train a deep learning model with smoothed noisy segmentations. Our proposed methods properly account for the uncertainty in tissue boundaries. We evaluated our method on 23 manually-segmented test images of a separate set of fetuses. Results show that our method achieves an average Dice similarity coefficient of 0.893 and 0.916 for the transient structures of younger and older fetuses, respectively. Our method generated results that were significantly more accurate than several state-of-the-art methods including nnU-Net that achieved the closest results to our method. Our trained model can serve as a valuable tool to enhance the accuracy and reproducibility of fetal brain analysis in MRI.

\end{abstract}


\vspace{12mm}

\section{Introduction}
\label{sec:introduction}

\subsection{Background and motivation}

\IEEEPARstart{F}{etal} magnetic resonance imaging (MRI) has emerged as an important and viable tool for assessing the development of brain in utero. It has enabled assessment of normal and abnormal brain growth trajectories in utero \cite{corbett20113d}. Moreover, fetal MRI may offer more accurate assessment and quantification of fetal brain development and degeneration when ultrasound images are inadequate \cite{hosny2010ultrafast,weisstanner2015mri}. Faster image acquisition methods \cite{yamashita1997mr} and superior super-resolution algorithms \cite{ebner2020automated,kainz2015fast,kuklisova2012reconstruction,gholipour2010robust} can now reconstruct high-quality 3D fetal brain images from stacks of 2D slices. These technical advancements have significantly improved the quality of fetal brain MRI. As a result, a growing number of works have successfully used fetal MRI to study various congenital brain disorders \cite{egana2013differences,mlczoch2013structural}. 
As the use of fetal MRI in clinical and research studies grows, quantitative image analysis methods become an urgent requirement. Automatic analysis methods can increase the speed, accuracy and reproducibility of quantification of fetal brain development. Accurate segmentation of the fetal brain into relevant tissue compartments is especially critical because many congenital brain disorders manifest themselves as changes in the size or shape of these tissues. Whereas manual segmentation is time-consuming and prone to high intra/inter-observer variability \cite{gousias2012magnetic}, automatic segmentation promises high speed and reproducibility. As a result, in recent years there have been multiple efforts to segment different tissue compartments in the fetal brain.

A review of automatic fetal and neonatal brain segmentation techniques in MRI can be found here \cite{makropoulos2018review}. Here, we focus primarily on deep learning (DL) methods. Deep learning-based segmentation of adult brain into different tissue compartments has been successfully attempted by several studies in recent years \cite{dolz20183d,sun20193d}. Overall, they show that DL methods are capable of accurately segmenting brain into relevant tissue compartments. Comparatively, much fewer studies have targeted the fetal brain tissue segmentation. For fetal cortical gray matter segmentation, one study proposed obtaining cheap annotations using an automatic segmentation method originally designed for neonatal brains \cite{fetit2020deep}. They used a human-in-the-loop method to refined those segmentations on selected 2D slices. A fully convolutional network (FCN) was trained using these annotations. Their method achieved an average Dice Similarity Score (DSC) of 0.76. Segmentation of cortical gray matter has been addressed by several other works. One work used a deep attentive FCN and reported a mean DSC of 0.87 \cite{dou2020deep}, whereas another study proposed integrating a topological constraint into the training loss function and achieved a mean DSC of 0.70 \cite{dumast2021segmentation}. Another study used the nnU-Net framework to segment the white matter, ventricles, and cerebellum in fetal brain MRI, achieving DSC values in the range 0.78-0.94 \cite{fidon2021distributionally}. To reduce the impacts of motion artifacts and partial volume effects, \cite{li2021cas} proposed a unified deep learning framework to jointly estimate a conditional atlas and predict a segmentation. The rationale for this approach is that the prior knowledge provided by the atlas can guide the segmentation where image quality is low. The idea of leveraging atlases to improve deep learning-based segmentation has been explored in several other works \cite{oguz2018combining, diniz2020esophagus, karimi2018prostate, zeng2018prostate, KARIMI2019186}. A succession of two FCNs was proposed by \cite{khalili2019automatic}, the first to extract the intracranial volume and the second to segment the brain tissue into seven compartments. This method achieved a mean DSC of 0.88. For segmenting the fetal brain into seven tissue compartments, another study used a single 2D UNet and achieved a mean DSC of 0.86 \cite{payette2020efficient}. \cite{payette2021automatic} compared a multi-atlas segmentation method with several DL methods for segmentation of fetal brain into seven tissue types and found that overall DL methods can achieve more accurate results.

\subsection{Segmentation with noisy labels}

Deep learning segmentation models, which represent the state of the art, require large accurately-labeled training datasets. Such datasets are especially difficult to come by in fetal MRI because the image quality is low and accurate multi-label segmentation of 3D images is very time-consuming. Despite recent progress in super-resolution reconstruction methods, 3D fetal MR images can suffer from residual motion and partial volume effects, making accurate delineation of tissue boundaries challenging and uncertain. Moreover, the fetal brain undergoes rapid and significant changes during the second and the third trimesters. Therefore, to develop an accurate DL model, training data should include a sufficiently large number of subjects at different gestational ages (GA) in order to fully capture the variability in the transient fetal brain structures.

Because detailed manual segmentation of a large number of 3D fetal brain images is impossible or prohibitive, an alternative strategy would be to use less accurate annotations. Scenarios with weak, partial, or noisy labels are very common in medical image analysis. Hence, training of DL models with imperfect labels has been the subject of intense research in recent years \cite{cheplygina2019, tajbakhsh2019, rajchl2016deepcut}. Such labels can often be obtained at low cost using automatic or semi-automatic methods. A review of the state of the art methods for handling the label noise in DL can be found in \cite{song2020learning}. We previously published a survey that is more focused on medical image analysis applications \cite{karimi2020deep}, where we identified six classes of methods for training DL models under strong label noise. Below, we describe two of the techniques that are more relevant to this work.

\textbf{Loss function.} There have been many efforts to devise loss functions that are tolerant to label noise \cite{zhang2018generalized, rusiecki2019trimmed}. These loss functions typically tend to down-weight the penalty on data samples that incur very high loss values, under the assumption that those data samples are likely to have wrong labels. Another group of loss functions and training procedures are based on estimating and incorporating a label transition matrix \cite{patrini2017making, sukhbaatar2014training}. Label transition matrix $T \in {\rm I\!R}^{L \times L}$, where $L$ is the number of labels, is meant to describe how the correct labels are flipped into incorrect labels. If we denote the clean and noisy class probability vectors with, respectively, $p_c \in {\rm I\!R}^{L}$ and $p_n \in {\rm I\!R}^{L}$, then we have $p_n= T p_c$. Hence, $T_{i,j}$ is the probability that the correct label $j$ is flipped to label $i$. Different approaches to estimating $T$ and using it in training DL classification models have been proposed in prior works \cite{thekumparampil2018robustness, bekker2016training}.

\textbf{Label smoothing.} Label smoothing has been extensively used in image classification \cite{pereyra2017regularizing} as well as in natural language processing applications \cite{chorowski2016towards}. However, very few studies have used label smoothing for segmentation applications. In fact, the standard label smoothing approach is unlikely to be suitable for segmentation applications. This is because, unlike classification where the whole image is represented with one probability vector, in segmentation a probability vector belongs to a single pixel/voxel and there are strong spatial correlations between the labels of nearby voxels. Standard label smoothing ignores those spatial correlations and essentially assumes that the probability that label $k$ is flipped to label $l \neq k$ is the same for all $l$, which is an unrealistic assumption. One study suggested smoothing the object boundaries in training data in order to improve the uncertainty calibration of the trained model \cite{islam2021spatially}. However, although they used the term ``spatially-varying'' to describe their method, they applied a fixed operation to all voxels. Another study proposed a label smoothing approach to improve the model uncertainty calibration for scene segmentation \cite{liu2021devil}. In the context of image colorization, one study used label smoothing to achieve more accurate scene segmentation \cite{nguyen2020image}. However, none of these studies have properly addressed the spatially-varying nature of boundary uncertainty in semantic segmentation.

Another challenge in fetal brain tissue segmentation is that it involves a large number of compartments that vary significantly in size. In this work, we aim to segment the fetal brain into more than 30 classes, where the volume of the smallest class is typically $10^4$ times smaller than the volume of the largest class. This can present a significant challenge for some of the loss functions that are commonly used to train DL segmentation models.

The goal of this work is to develop methods for accurate fetal brain tissue segmentation in MRI. Most prior works have segmented only a single tissue (e.g., \cite{dou2020deep,fetit2020deep}) or have divided the fetal brain into a small number of tissue compartments (e.g., \cite{payette2020efficient,fidon2021distributionally}). In this work, we consider more than 30 relevant and important tissue compartments. To capture the rapid brain growth in utero and the complex developmental trajectories of these tissues, we use a training dataset of 272 images covering the gestational age between 19 and 39 weeks. Instead of manually annotating this dataset, which would have been prohibitive, we use a combination of automatic atlas-based segmentation and manual correction of gross errors. Using the expert-estimated boundary uncertainty for different tissues, we develop novel methods for label smoothing and for training a DL model with the smoothed labels. We evaluate our trained model and compare it with several alternative methods on a set of manually-segmented test images. We show that our method achieves high segmentation accuracy and outperforms several state of the art methods.

\section{Materials and methods}

\subsection{Data and annotation procedures}
\label{data_and_annotation}

Data from 294 fetuses with GA between 19.6 and 38.9 weeks (mean 30.6; standard deviation 5.3) were used in this study. These data were collected in studies approved by the institutional review board committee. Written informed consent was obtained from pregnant women volunteers who participated in research MRI scans for fetal MRI. All images were collected with 3-Tesla Siemens Skyra, Trio, or Prisma scanners using 18 or 30-channel body matrix coils via repeated T2-weighted half-Fourier acquisition single shot fast spin echo (T2wSSFSE) scans in the orthogonal planes of the fetal brain. The slice thickness was 2mm with no inter-slice gap, in-plane resolution was between 0.9mm and 1.1mm, and acquisition matrix size was $256 \times 204$, $256 \times 256$, or $320 \times 320$. Volumetric images were reconstructed using an iterative slice-to-volume reconstruction algorithm \cite{kainz2015fast}, brain extracted and registered to a standard atlas space in a procedure described in~\cite{gholipour2017normative}. The resulting 3D images had isotropic voxels of size 0.8mm. We selected 22 of these fetuses and set them aside as ``test subjects'' for final evaluation. The test subjects had GA between 23.3 and 38 weeks (mean 32.9; standard deviation 4.14). We used the remaining, completely independent, 272 subjects as ``training subjects'' to develop/train our methods and also to train the competing techniques.

We manually segmented the test images in detail. To speed up the process, we first generated automatic segmentation for each subject with a multi-atlas segmentation method using a publicly available atlas \cite{gholipour2017normative}. This is a four-dimensional (i.e., spatio-temporal) atlas that covers the GA range between 19 and 39 weeks at one-week intervals. For each test fetus, we registered atlases that were within one week GA of the fetus using a diffeomorphic deformable registration algorithm. We then used the probabilistic Simultaneous Truth and Performance Level Estimation (STAPLE) algorithm \cite{akhondi2013simultaneous} to fuse the segmentations. Then, experienced annotators carefully refined all labels in several rounds until the segmentations were consistent and free of any non-trivial errors. This was a laborious effort, which required 4-10 days of work for each scan. We used these manual segmentations as ``ground truth'' to test our method and competing techniques.

We then segmented the 272 training scans using a similar two-step approach, but with one major difference. Specifically, in the manual refinement step the annotators only corrected major errors, which on average required approximately two hours of work for each scan that had major errors. This was done because manually segmenting all 272 images with the same level of detail as done for the test images would have been impossible given the annotators' time. Furthermore, in order to account for the potential errors and uncertainties in these segmentations, we asked the annotators to specify the degree of uncertainty in the boundary of each tissue type. Given that all images had the same spatial resolution, this uncertainty was expressed in terms of the number of voxels. The boundary uncertainty specified by the annotators varied considerably for different tissues. For example, for lateral ventricle the boundary uncertainty was 0 voxels, meaning that the boundary was generally unambiguous, while for the caudate nuclei it was 2 voxels. 

Labels considered in this study included the following: hippocampus (HP)\textsuperscript{$\dagger$}, amygdala (AM)\textsuperscript{$\dagger$}, caudate nuclei (CD)\textsuperscript{$\dagger$}, lentiform nuclei (LN)\textsuperscript{$\dagger$}, thalami (TH)\textsuperscript{$\dagger$}, corpus callosum (CC), lateral ventricles (LV)\textsuperscript{$\dagger$}, brainstem (ST), cerebellum (CR)\textsuperscript{$\dagger$}, subthalamic nuclei (SN)\textsuperscript{$\dagger$}, hippocampal commissure (HC), fornix (FN), cortical plate (CP)\textsuperscript{$\dagger$}, subplate zone (SP)\textsuperscript{$\dagger$}, intermediate zone (IZ)\textsuperscript{$\dagger$}, ventricular zone (VZ)\textsuperscript{$\dagger$}, white matter (WM)\textsuperscript{$\dagger$}, internal capsule (IC)\textsuperscript{$\dagger$}, CSF, and ganglionic eminence (GE)\textsuperscript{$\dagger$}. A $\dagger$ next to a label in this list indicates that separate components in the left and the right brain hemispheres were considered for that tissue type. In the rest of this paper, we use the acronyms defined above to refer to these tissues and structures. Following the construction of the atlas \cite{gholipour2017normative}, there is one age-dependent difference in tissue labels. Specifically, fetuses that are younger than 32 weeks GA have separate SP and IZ labels, whereas for fetuses that are 32 weeks GA and older these two tissue types are merged as a single label: WM. As a result, younger fetuses have 33 tissue labels, whereas older fetuses have 31 labels (in addition to the background label). Therefore, for our method and also for all competing methods, we trained two separate models for the two fetal age groups.

\subsection{Development of a DL-based segmentation method}

\subsubsection{Label smoothing.}

In order to account for the inherent and unavoidable uncertainty in the tissue boundaries, we propose a spatially-varying label smoothing method. Our label smoothing method is presented in Algorithm 1. It is based on the fact that label uncertainty is limited to tissue boundaries and depends on the tissues that meet at the boundary. Specifically, our label smoothing strategy is based on the observation that, as confirmed by our annotators, the boundary uncertainty is dictated by the \emph{more certain} tissue. For example, consider a boundary where one of the adjoining tissues has an uncertainty of two voxels (less certain) but the other has an uncertainty of zero voxels (more certain). The boundary will have an uncertainty of zero because the tissue with more certain boundary resolves the ambiguity.

Let us denote the expert-provided semi-automatic segmentation with $Y \in {\rm I\!R}^{N \times L}$, where $N$ is the number of voxels and $L$ is the number of labels. Note that we can also write $Y \in \{ 0,1 \}^{N \times L}$ because $Y$ is a hard (0 or 1) label. For each voxel, $Y$ is a one-hot probability vector $\textbf{e}_k$ (which equals 1 at location $k$ and 0 elsewhere), where $k$ is the indicated tissue label for that voxel. We define $Y^* \in {\rm I\!R}^{N}$ as $Y^*=\argmax_{l \in [1,L]} (Y)$; in other words $Y(i)= \textbf{e}_k  \Rightarrow Y^*(i)= k$. We denote with $U \in {\rm I\!R}^{N}$ the tissue boundary uncertainty map. $U$ is obtained from $Y$ by simply setting $U(i)$ to the boundary uncertainty of the tissue label for $Y(i)$. If, for example, tissue label for $Y(i)$ is caudate nuclei (CD), then $U(i)=2$ because the boundary uncertainty for CD is two voxels. We use $i^r$ to denote all voxels that are within a distance $r$ from voxel $i$; in other words $i^r= \{k , \| k-i \| \leq r \}$. Throughout this paper, $\| . \|$ denotes the $\ell_2$-norm. Also note that all voxel indices are in fact 3D indices (i.e., they have $xyz$ elements) and, hence, $\| k-i \|$ is a distance in ${\rm I\!R}^3$. However, we use single letters for indices in order to simplify the notation. Finally, we use $Y(i^r)$ to denote the ``patch'' of $Y$ centered on voxel $i$ with a radius $r$. 

\begin{algorithm}
\caption{The proposed segmentation label smoothing algorithm.}\label{alg:two}
\textbf{Input:} hard segmentation labels $Y \in {\rm I\!R}^{N \times L}$,\\
\hspace{10mm} tissue boundary uncertainty map $U \in {\rm I\!R}^{N}$,\\ \hspace{10mm} and upper bound on tissue boundary \\ \hspace{10mm} uncertainty $R$. \\
\textbf{Output:} smoothed segmentation labels $Y_S \in {\rm I\!R}^{N \times L}$.\\
\textbf{Initialize:} $Y_S = \bm{0} \in {\rm I\!R}^{N \times L}$.\\
\For{$i \in [1,N]$}{
    \eIf{$\text{std} [ Y^*(i^R) ]= 0$ or $\text{min} [ U(i^R)]=0$}{
    $Y_S(i)= Y(i)$;
  }{
$r_u= \text{min} [ U(i^R)]$; \\
$W_k \propto  \exp( - \| k-i \| / r_u ) \hspace{3mm} \forall k \in i^{r_u}$; \\
$Y_S(i)[l]= \sum_{i^{r_u}} W \odot \mathcal{P} \big( Y(i^{r_u})=l \big) $ \\ $ \hspace{43mm} \forall l \in [1,L] $; \\
  }
}
\end{algorithm}

\begin{figure*}[!ht]
\includegraphics[width=\textwidth]{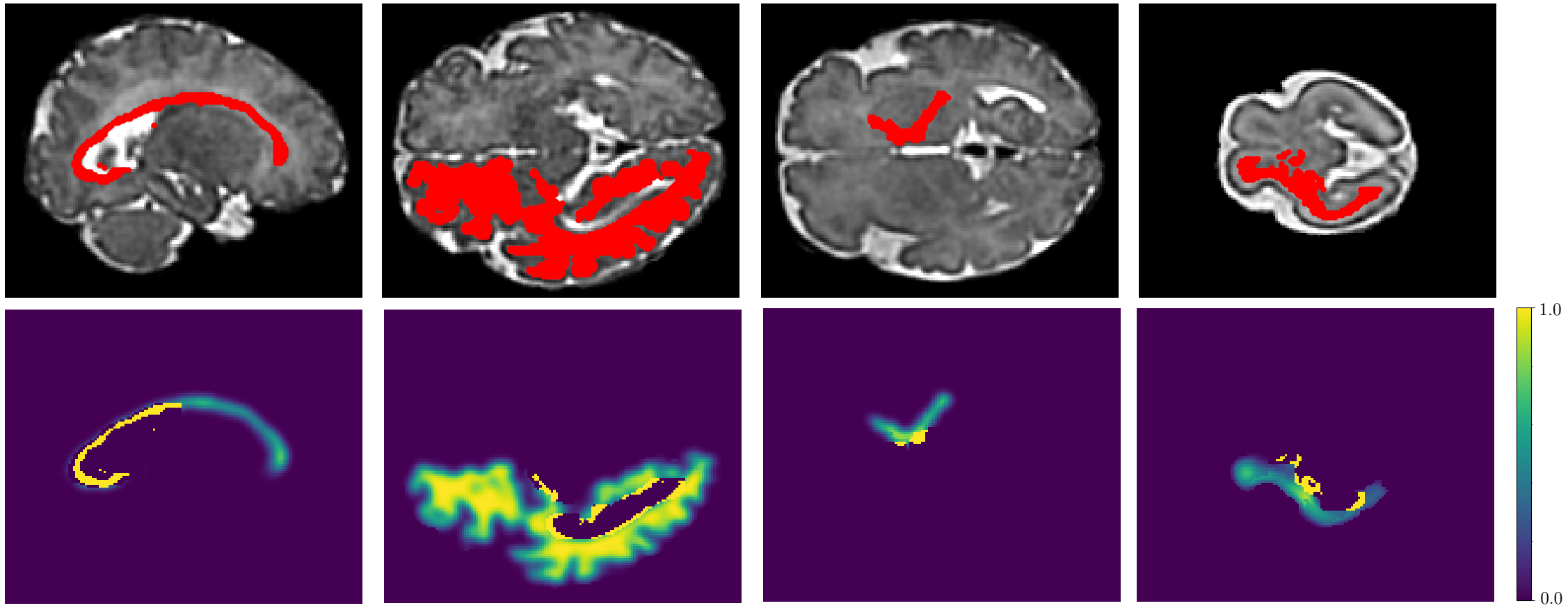}
\caption{Examples of label smoothing performed by our proposed method (Algorithm 1). Note that for each image our algorithm performs the smoothing simultaneously on all labels, but here we show a single label in each example for better visualization. In each of the examples presented here, the top image shows the original label obtained via multi-atlas segmentation followed by manual correction of large errors, super-imposed on the T2 image. The lower images show the smoothed label.}
\label{label_smoothing}
\end{figure*}

\vspace{6mm}

Now, given manual segmentation labels, $Y \in {\rm I\!R}^{N \times L}$, we would like to compute smoothed labels, $Y_S \in {\rm I\!R}^{N \times L}$, that account for uncertain tissue boundaries. To do this for voxel $i$, we first consider $i^R$, where $R=4$ is the upper bound of boundary uncertainty reported by our annotators for all tissues. If $Y(i^R)$ is homogeneous, that is, $\text{std} [ Y^*(i^R) ]= 0$, it means that voxel $i$ is far from tissue boundaries since all voxels in $i^R$ have the same label. Otherwise, voxel $i$ is close to a boundary. In that case, we compute the boundary uncertainty as $r_u= \text{min} [ U(i^R)]$, i.e., the minimum of the uncertainty of the tissues in $i^R$. This is done following the justification provided above since the tissue with the lower uncertainty determines the uncertainty of the boundary. We then use a weighted average of tissue probabilities in $i^{r_u}$ to compute the smoothed class probability vector for this voxel. Specifically, we use a weight matrix $W_k \propto  \exp( - \| k-i \| / \tau ) \hspace{3mm} \forall k \in i^{r_u}$, which gives higher weights to voxels that are closer to voxel $i$. We have found that the kernel width, $\tau$, should depend on the patch size, which is in turn related to the boundary uncertainty. This also makes intuitive sense because for more uncertain boundaries (larger $r_u$) a larger kernel width should be used to achieve a higher degree of smoothing. In the experiments reported in this paper we simply set $\tau= r_u$, which we found empirically to work well. In order to ensure that the class probability vector for each voxel sums to one, $W_k$ is normalized to sum to one. For each class label $l$, the smoothed probability at voxel $i$ is then computed as:

\begin{equation}
Y_S(i)[l]= \sum W \odot \mathcal{P} \big( Y(i^{r_u})=l \big),
\end{equation}

\noindent
where $\odot$ denotes element-wise (Hadamard) product and the summation is carried out over all voxels in the patch $i^{r_u}$. Figure \ref{label_smoothing} shows example label smoothing results generated with our proposed method. These examples show that our method smooths the boundary of each tissue/label based not only on the boundary uncertainty of that tissue but also on the boundary uncertainty of the adjoining tissues. For example, at the locations where a tissue shares a boundary with the lateral ventricles (which have a boundary uncertainty of zero voxels), the boundary becomes unambiguous and no smoothing is performed.

\subsubsection{Loss function.}

We use a loss function that treats certain and uncertain regions differently. We use $M$ to denote a binary mask that shows voxels with uncertain (smoothed) labels. $M$ is easily obtained as voxels where the maximum class probability in $Y_s$ is not equal to one, or equivalently, as voxels whose labels are altered in the process of label smoothing. In other words:

\begin{equation}
\Big( Y_s[i] \neq Y[i] \Big) \Rightarrow M[i]=1. 
\end{equation}

Then, our loss function is:

\begin{equation}
Loss(\hat{Y},Y_S) = \sum_{M=0} \mathcal{L}(\hat{Y},Y_S) + \sum_{M=1} \sum_{l} T^{-\text{T}} \mathcal{L}( \hat{Y}, Y_S ),
\end{equation}

\noindent where $T^{-\text{T}}$ is the inverse of transpose of $T$, $\hat{Y}$ is the segmentation map predicted by our DL model, and $\mathcal{L}$ is the base loss function, which we choose to be the cross-entropy. It has been shown that multiplication with $T^{-\text{T}}$, for the uncertain boundary voxels in the second loss term above, leads to an unbiased minimizer \cite{patrini2017making}. In other words, the minimizer of the corrected loss function is the same as the minimizer obtained with clean (true) labels. Even though in this work $T^{\text{T}}$ was non-singular, instead of $T^{-\text{T}}$ we used $(T^{\text{T}}+ \lambda I)^{-1}$ with $\lambda=1$ as suggested in prior works \cite{patrini2017making} because it resulted in faster and more stable training.

\begin{figure*}[!ht]
\centering
\includegraphics[width=115mm]{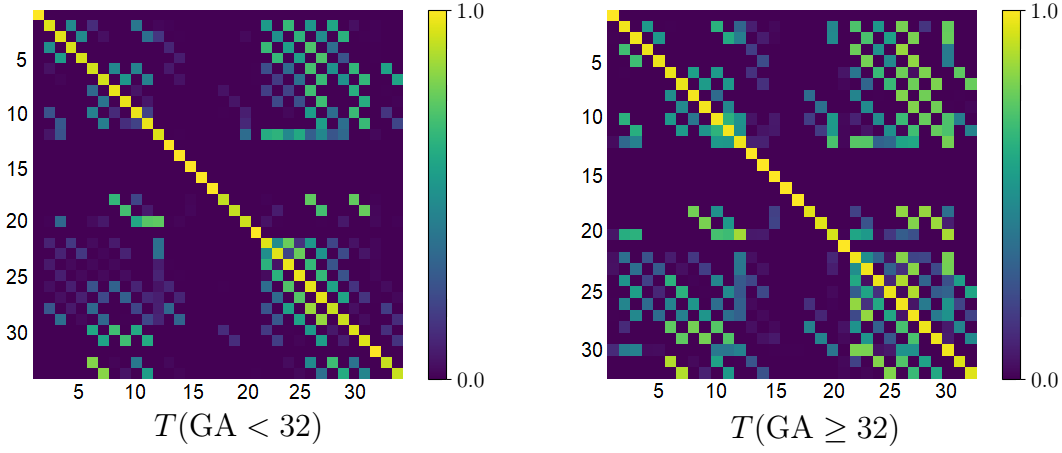}
\caption{Label transition matrices, $T$, for the younger (left) and older (right) age groups. We have used a log transformation, $\log (T+0.001)$, to display these matrices in order to better highlight the smaller off-diagonal elements.   Label numbers have been shown on the rows and columns of the matrices. For younger fetuses the label names are as follows (acronyms have been defined in the text in Section \ref{data_and_annotation}). 1: background, 2: HP left, 3: HP right, 4: AM left, 5: AM right, 6: CD left, 7: CD right, 8: LN left, 9: LN right, 10: TH left, 11: TH right, 12: CC, 13: LV left, 14: LV right, 15: ST, 16: CR left, 17: CR right, 18: SN left, 19: SN right, 20: HC, 21: FN, 22: CP left, 23: CP right, 24: SP left, 25: SP right, 26: IZ left, 27: IZ right, 28: VZ left, 29: VZ right, 30: IC left, 31: IC right, 32: CSF, 33: GE left, 34: GE right. For older fetuses, the first 23 labels are the same as those for younger fetuses, and the remaining labels are as follows. 24: VZ left, 25: VZ right, 26: WM left, 27: WM right, 28: IC left, 29: IC right, 30: CSF, 31: GE left, 32: GE right.}
\label{T_matrix}
\end{figure*}

We computed the label transition matrix $T$ empirically from our training data. Specifically, after applying Algorithm 1 on the training labels, we computed $T$ from 50 of the training images as $T_{i,j}= \Sigma_{ \{ Y=j \cup M=1 \} } \mathcal{P}(Y_s=i)$, where summation is performed over all images. This empirical estimation is based on the standard definition of the label transition matrix. We normalized each column of $T$ to sum to unity because it should be a left stochastic matrix. Figure \ref{T_matrix} shows our estimated $T$ for younger and older fetuses.

\subsubsection{Implementation and experiments.}

As the baseline for implementation and comparison of different methods (described below) we used the nnU-Net framework \cite{isensee2021nnu}. nnU-Net is considered to be the state of the art in medical image segmentation. In particular, on 53 different segmentation tasks, nnU-Net has shown that with proper selection of the training pipeline settings, standard 3D U-Net architectures \cite{cciccek20163d} can match or outperform more elaborate network architectures. Hence, we adopt nnU-Net's network architecture (i.e., a 3D U-Net) and follow its training and inference strategies. We assess the effectiveness of our proposed methods by comparing against the methods below.

\begin{itemize}

\item \textbf{nnU-Net.} We followed the methods and settings in \cite{isensee2021nnu}. In particular, we used the default loss function, which is the sum of cross-entropy and Dice.

\item We tried three alternative loss functions. These included 1) \textbf{Generalized Dice} \cite{sudre2017generalised} which has been proposed to improve the segmentation accuracy when the target objects are small or suffer from severe class imbalance, 2) \textbf{Focal loss} \cite{lin2017focal}, which has been devised to address the extreme class imbalance by down-weighting the impact of structures that are easier to segment, and 3) improved Mean Absolute Error (\textbf{iMAE}) \cite{wang2019derivative}, which has been proposed for training DL models under strong label noise.

\item \textbf{Training on clean labels}. In this approach, instead of training on the 272 images with noisy segmentations, we used leave-one-out cross-validation to train and test using the 22 images with highly accurate labels, which we have called ``test subjects'' so far. Of those 22 subjects, 8 were younger fetuses ($\text{GA}<32$) and 14 were older fetuses ($\text{GA} \geq 32$). Therefore, for younger fetuses each time we trained the model on 7 images and tested on the remaining image, and for older fetuses each time we trained on 13 of the images and tested on the remaining image. In this approach, each of the 22 images was used as a test image in exactly one of the experiments. Therefore, the test set for this experiment was the same as for the other methods. Since the number of training images in this approach is small, we performed this experiment both without and with transfer learning, which is a common method for dealing with limited training data \cite{cheplygina2019,karimi2021transfer}. For transfer learning, we used 400 subjects from the Developing Human Connectome Project dataset (DHCP) \cite{hughes2017dedicated,cordero2016sensitivity} for pre-training. To the best of our knowledge, this is the most similar public dataset for the application considered in this work. It includes T2 images and tissue segmentation maps with 87 labels for newborns with GA in the range 29-45 weeks. For transfer learning, we pre-trained the network using this dataset. We then replaced the last network layer (i.e., the segmentation head) with a new head to match the number of labels considered in this work (34 for the younger fetuses and 32 for the older fetuses). We fine-tuned the pre-trained network on our data following the same leave-one-out cross-validation approach described above. We performed this separately for younger and older fetuses. Moreover, as in all experiments except for UNet++ and DeepLab mentioned below, the network architecture was the same 3D U-Net from the nnU-Net framework.

\item \textbf{Standard label smoothing}. Following the standard label smoothing approach \cite{szegedy2016rethinking, pereyra2017regularizing, muller2019does}, we set $Y_s[i]= (1-\alpha) Y[i] + \alpha / L$ for every voxel, except for the background voxels. We use $\alpha= 0.1$, which previous studies have shown to be a good setting \cite{pereyra2017regularizing, muller2019does}.

\item \textbf{SVLS} \cite{islam2021spatially}. SVLS is a label smoothing method, recently proposed for medical image segmentation.

\item \textbf{UNet++.} In order to also investigate the potential impact of network architecture, we compared with UNet++ \cite{zhou2018unet}. This is a more elaborate nested U-Net architecture that has been proposed specifically for medical image segmentation, which claims to be better than the standard U-Net.

\item \textbf{DeepLab.} DeepLab is a popular deep learning model for semantic segmentation \cite{chen2017deeplab}. The novelty of the network architecture is the use of atrous convolutions. Furthermore, a Conditional Random Field is used to improve the resolution of the segmentation predictions.

\end{itemize}

\begin{table*}[!ht]
 \caption{Segmentation accuracy metrics presented separately for younger and older fetuses. The metrics were computed separately for each label; this table presents mean $\pm$ standard deviation over all labels. Best results for each metric are in bold. We used paired t-tests to compare our proposed method with every other method. Asterisks in this table denote significantly better results for the proposed method than all other methods} (at a significant threshold of $p<0.001$). In the Method column in this table, T.L. stands for transfer learning.
\label{table:results}
   \begin{center}
    \begin{tabular}{ L{25mm} L{55mm} C{25mm} C{25mm} C{25mm} }
\hline
Dataset & Method & DSC & HD95 (mm) & ASSD (mm)   \\ \hline
\multirow{11}{*}{Younger fetuses} & nnU-Net & $0.872 \pm 0.063$  & $0.99 \pm 0.11$ & $0.26 \pm 0.14$  \\ 
& Generalized Dice & $0.845 \pm 0.087$  & $1.09 \pm 0.12$ & $0.32 \pm 0.26$   \\ 
& Focal loss       & $0.839 \pm 0.080$  & $1.15 \pm 1.20$ & $0.30 \pm 0.19$   \\ 
& iMAE             & $0.865 \pm 0.075$  & $1.06 \pm 0.15$ & $0.26 \pm 0.17$   \\ 
& Training on clean labels (without T.L.) & $0.863 \pm 0.068$  & $1.09 \pm 0.14$ & $0.28 \pm 0.16$   \\ 
& Training on clean labels (with T.L.) & $0.866 \pm 0.062$  & $1.03 \pm 0.13$ & $0.27 \pm 0.17$   \\ 
& Standard label smoothing & $0.833 \pm 0.084$  & $1.08 \pm 0.17$ & $0.34 \pm 0.21$   \\ 
& SVLS & $0.843 \pm 0.074$  & $1.07 \pm 0.17$ & $0.30 \pm 0.18$   \\ 
& DeepLab & $0.851 \pm 0.072$  & $1.11 \pm 0.15$ & $0.30 \pm 0.15$   \\ 
& UNet++ & $0.866 \pm 0.060$  & $1.02 \pm 0.14$ & $0.27 \pm 0.15$   \\ 
& Proposed method & $\bm{0.893 \pm 0.066^*}$  & $\bm{0.94 \pm 0.13}^*$ & $\bm{0.23 \pm 0.13}^*$   \\ \hline
\multirow{11}{*}{Older fetuses} & nnU-Net & $0.896 \pm 0.066$  & $0.98 \pm 0.11$ & $0.36 \pm 0.12$   \\
& Generalized Dice &      $0.866 \pm 0.070$  & $1.16 \pm 0.11$ & $0.46 \pm 0.15$   \\ 
& Focal loss &            $0.861 \pm 0.068$  & $1.16 \pm 0.16$ & $0.42 \pm 0.16$   \\ 
& iMAE                  & $0.880 \pm 0.064$  & $1.09 \pm 0.17$ & $0.41 \pm 0.20$   \\ 
& Training on clean labels (without T.L.) & $0.877 \pm 0.073$  & $1.12 \pm 0.14$ & $0.40 \pm 0.18$   \\ 
& Training on clean labels (with T.L.) & $0.880 \pm 0.070$  & $1.04 \pm 0.15$ & $0.40 \pm 0.20$   \\ 
& Standard label smoothing & $0.853 \pm 0.071$  & $1.16 \pm 0.12$ & $0.39 \pm 0.23$   \\ 
& SVLS & $0.856 \pm 0.077$  & $1.10 \pm 0.13$ & $0.37 \pm 0.27$   \\ 
& DeepLab & $0.865 \pm 0.074$ & $1.21 \pm 0.19$ & $0.43 \pm 0.26$   \\ & UNet++ &                   $0.885 \pm 0.070$  & $1.08 \pm 0.16$ & $0.38 \pm 0.23$   \\ 
& Proposed method & $\bm{0.916 \pm 0.059}^*$  & $\bm{0.94 \pm 0.13}^*$ & $\bm{0.25 \pm 0.09}^*$   \\ \hline
\end{tabular}
  \end{center}
\end{table*}

Note that in all of the above approaches, except for UNet++ and DeepLab, we followed the same nnU-Net framework for the choice of network architecture and training settings. We refer to \cite{isensee2021nnu} for the details of this framework. For UNet++, we followed the settings of the original paper \cite{zhou2018unet}. Furthermore, as mentioned above, for all methods except for ``Training on clean labels'', we had 272 training images. For each method, we first selected a good initial learning rate using a subset of 100 images. We then trained the model using the selected initial learning rate on all 272 images. We used the ``poly'' learning rate decay as in \cite{isensee2021nnu}. All training and test runs were performed using TensorFlow 1.14 under Python 3.7 on a Linux computer with an NVIDIA GeForce GTX 1080 GPU. The source code, trained model, and sample image data and segmentation labels for this work have been made publicly available at \url{https://github.com/bchimagine/fetal_tissue_segmentation}.

\section{Results and Discussion}

Table \ref{table:results} shows the summary of the segmentation accuracy results in terms of Dice Similarity Coefficient (DSC), 95 percentile of the Hausdorff Distance (HD95), and Average Symmetric Surface Distance (ASSD). Our method has achieved the best results in terms of all three criteria. To determine the statistical significance of the differences, we performed paired t-tests to compare our method with every competing method in terms of these three criteria. These tests showed that, with a $p$-value threshold of 0.001, our method achieved significantly higher DSC and significantly lower HD95 and ASSD than all other methods, both for younger and older age groups.

In terms of all metrics, nnU-Net was the second best method after our proposed method. As we mentioned above, we used the default loss function, which is the sum of Dice and cross-entropy. Compared with this default loss function, Generalized Dice and Focal Loss performed poorly because they systematically missed one or two of the structures. That is, with Generalized Dice and Focal Loss, the network output for one or two of the labels was empty. The missed structure(s) changed with network weight initialization, but they were usually the smaller structures such as amygdala, caudate, or subthalamic nuclei. Overall, in this application with more than 30 labels we have found that loss functions based on Dice do not perform well. This may be due to the fact that the overall loss is the sum of the loss on individual labels and as the number of labels increases the relative contribution of each of the labels to the total loss becomes smaller. Since the Dice is limited to the range [0,1], the worst-case effect of a label on the total loss is ($1/L$), where $L$ is the number of labels. In our application with $L \approx 33$, the effect of completely missing one of the labels is only $3\%$. As a result, training is prone to ignoring one the labels entirely and proceed to reduce the overall loss by improving the segmentation of the other labels. 

\begin{figure*}[!ht]
\centering
\includegraphics[width=150mm]{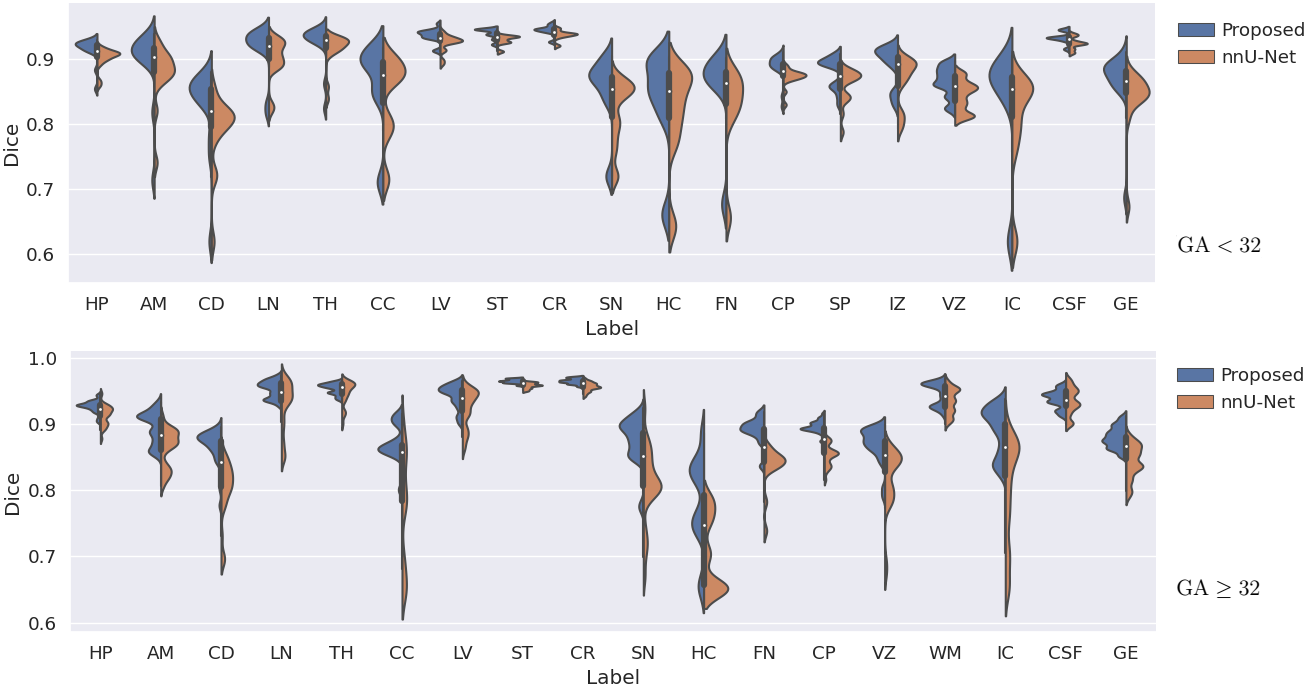}
\caption{Comparison of our method and nnU-Net in terms of DSC for different structures on younger (top) and older (bottom) fetuses.}
\label{proposed_vs_unet_DSC}
\end{figure*}

Compared with the Generalized Dice and the Focal Loss, the iMAE loss performed comparatively better and always segmented all the labels. However, it did not perform as well as nnU-Net's default loss. The iMAE loss has been proposed to strike a balance between the Mean Absolute Error (MAE) and cross-entropy. Although some studies in image classification have shown improved accuracy with MAE and iMAE when label noise is high, our results show that in the application considered in this study they do not lead to the highest segmentation accuracy. Another limitation of the iMAE loss is much longer training time, as we discuss further below.

\begin{figure*}[!ht]
\centering
\includegraphics[width=150mm]{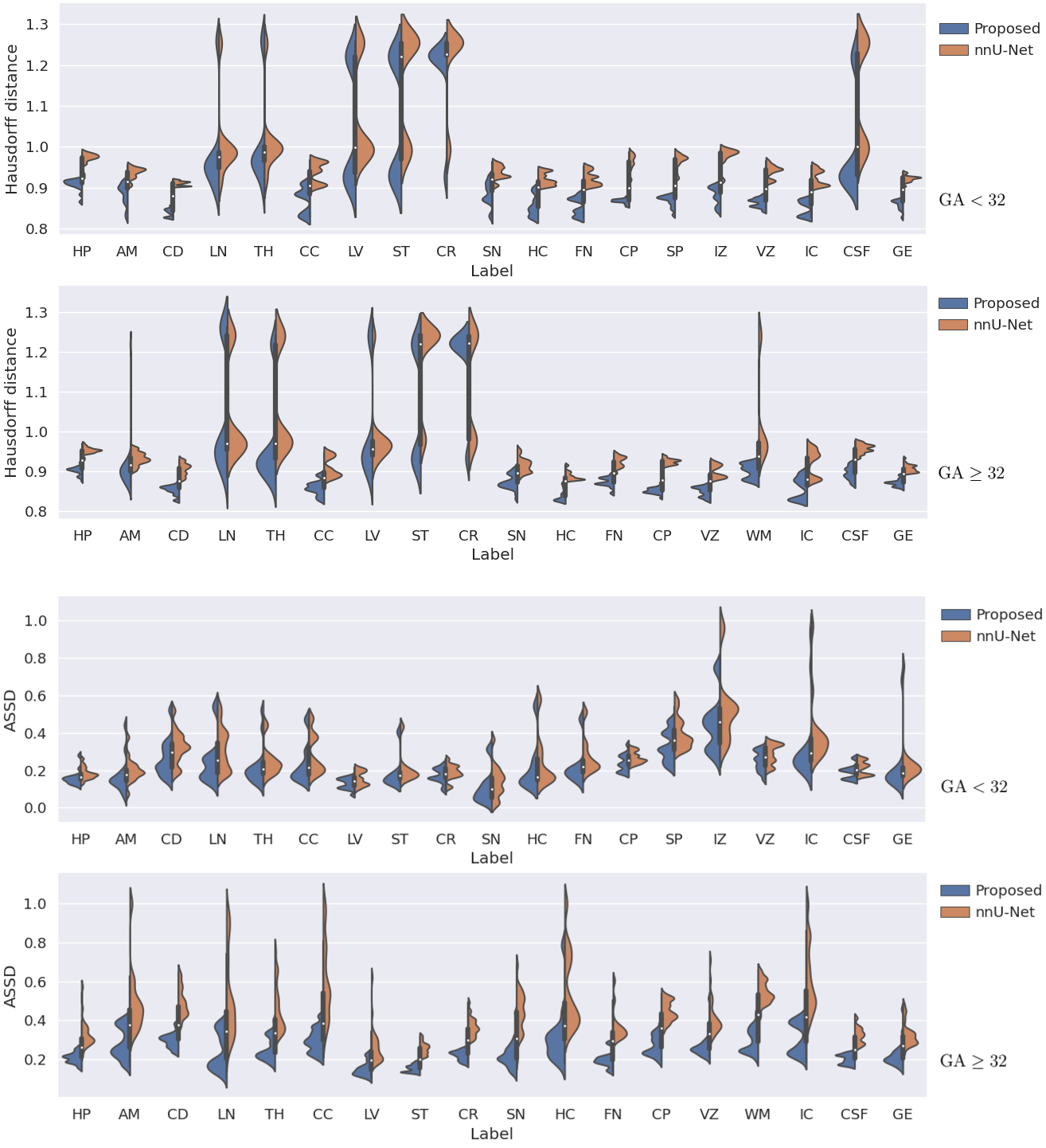}
\caption{Comparison of our method and nnU-Net in terms of HD95 (the top two plots) and ASSD (the bottom two plots) on younger and older fetuses. The GA group for each plot is shown on the right of that plot.}
\label{proposed_vs_unet_distance}
\end{figure*}

\begin{figure*}[!ht]
\centering
\includegraphics[width=0.90\textwidth]{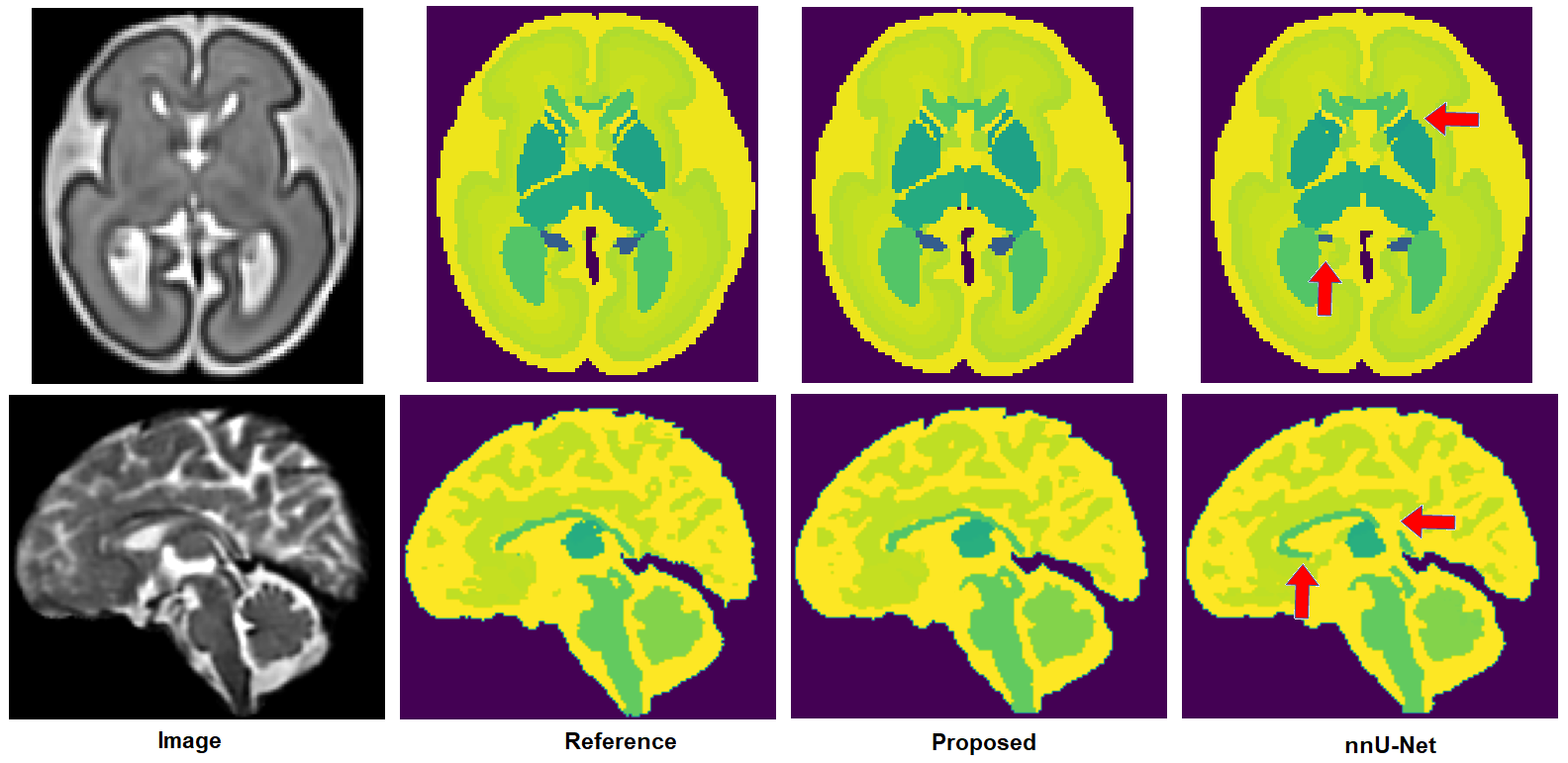}
\caption{Two example segmentation maps predicted by our proposed method and nnU-Net. Red arrows point to some of nnU-Net's segmentation errors. The top example is for a younger fetus (GA= 26.7 weeks), whereas the bottom example is for an older fetus (GA= 38 weeks).}
\label{visualize_full}
\end{figure*}

Training on clean labels was not effective, evidently because of the much reduced number of training images. Although this approach used more accurate manual labels for training, it was limited to far fewer training images (7 or 13 training in leave-one-out cross-validation experiments, compared with 272 training images for the other methods in Table \ref{table:results}). Although in some applications 7-13 training images may be adequate to achieve high segmentation accuracy \cite{karimi2021transfer}, the application considered in this work is especially challenging due to the rapid fetal brain development and significant changes in the brain size and shape. Because of the rapid developments in the shape and complexity of structures such as the cortical gray matter, much larger numbers of training images are needed to allow the network to effectively learn these structures across the gestational age. As mentioned in the Methods section above, this experiment was performed both without and with transfer learning. As shown in Table \ref{table:results}, there was a consistent but small improvement in segmentation accuracy due to transfer learning. We used paired t-tests to assess the statistical significance of these differences. The tests showed a significant reduction in HD95 ($p<0.001$) for both younger and older fetuses, although no significant differences ($p \approx 0.16-0.35$) in DSC or ASSD were found for either younger or older fetuses. The results presented for transfer learning in Table \ref{table:results} were obtained by fine-tuning all layers of the pre-trained network. We experimented with other transfer learning approaches such as shallow fine-tuning \cite{tajbakhsh2016, karimi2021transfer} but did not achieve better results. The results of these experiments suggest that, in the application considered in this work, training with a small number of manually labeled images cannot achieve the same level of segmentation accuracy as training with a much larger number of training images with less accurate labels.

Standard label smoothing and SVLS did not work well and achieved worse accuracy metrics than nnU-Net. Standard label smoothing treats all voxels, including voxels that are far from any tissue boundary, in the same way. The SVLS method only smooths the boundary voxels, but it uses a label smoothing approach that does not take into account the actual tissue-dependent boundary uncertainties. The results obtained with these two methods shows that un-informed or spatially uniform label smoothing is not useful for segmentation with noisy labels. Finally, UNet++ and DeepLab did not perform as well as nnU-Net, confirming the arguments presented by \cite{isensee2021nnu} that more elaborate network architectures and post-processing operations do not necessarily lead to better results.

\begin{figure*}[!ht]
\centering
\includegraphics[width=0.92\textwidth]{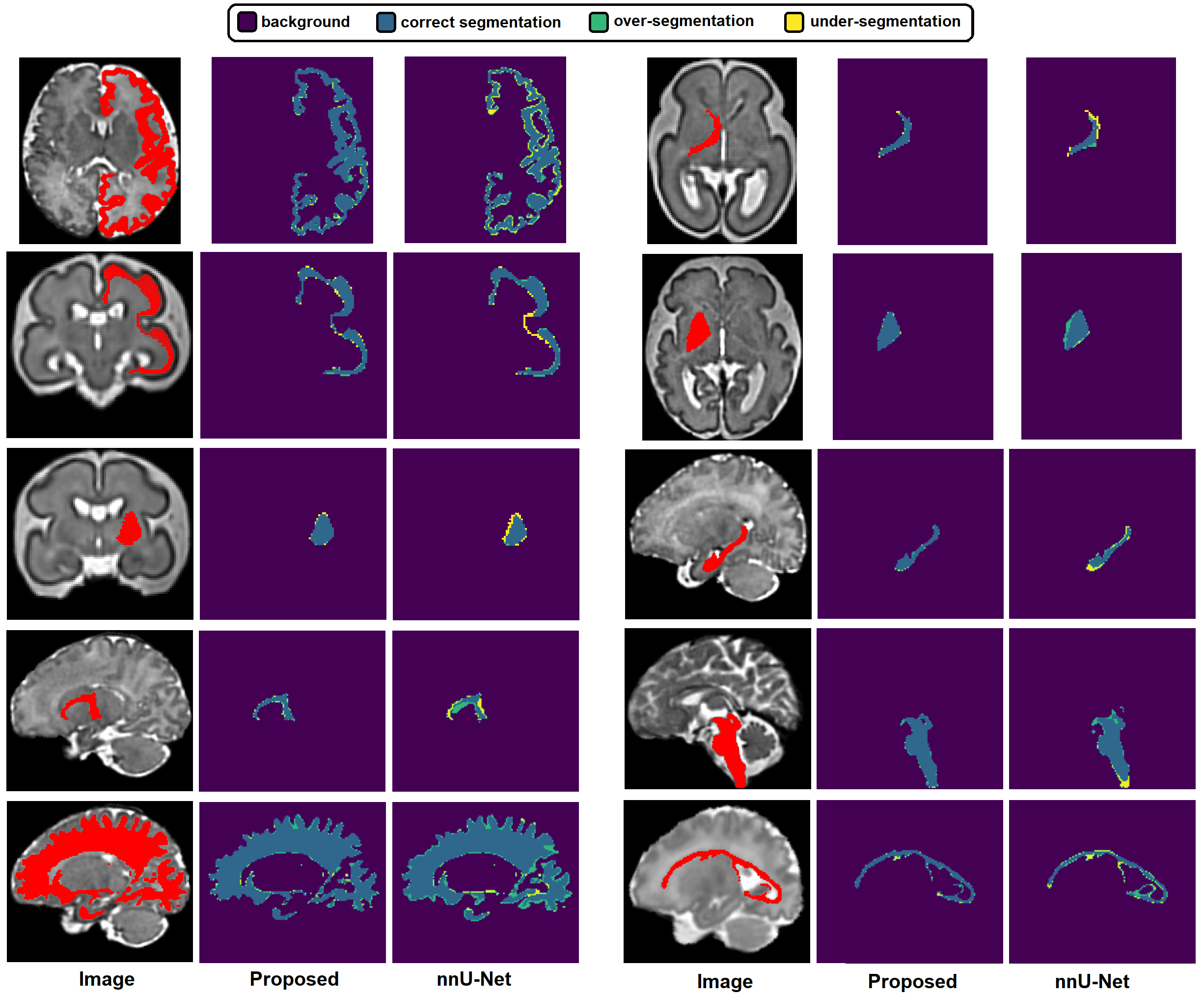}
\caption{Example segmentations predicted by our proposed method and nnU-Net. Each of the ten examples shows one isolated structure and highlights correct segmentations, over-segmentations and under-segmentations.}
\label{visualize_single}
\end{figure*}

Figure \ref{proposed_vs_unet_DSC} shows more detailed label-specific DSC comparison of our method with nnU-Net, which was better than the other competing methods in terms of the overall segmentation accuracy, as shown in Table \ref{table:results}. For structures that have separate left and right labels, we have combined the two labels into one label in order to produce a less cluttered plot. On younger fetuses, our method achieved significantly higher DSC than nnU-Net on all structures ($p<0.001$), except for CC, where the difference was not significant. For older fetuses, our method achieved significantly higher DSC on 14 of the structures ($p<0.001$), while on the remaining four structures (LN, TH, WM, and CSF) although the mean DSC for our method was higher, the differences were not significant. Figure \ref{proposed_vs_unet_distance} shows similar plots for HD95 and ASSD. For HD95, paired t-tests showed that our proposed method achieved significantly ($p<0.001$) lower errors than nnU-Net on all structures except for LN and CR on older fetuses. For ASSD, our method achieved significantly ($p<0.001$) lower errors on all structures except for CR on older fetuses.  

Figure \ref{visualize_full} shows two example segmentation maps, for one younger fetus and one older fetus, predicted by our method and nnU-Net. Both our proposed method and nnU-Net succeed in segmenting different structures with good accuracy. Nonetheless, the segmentations produced by the proposed method were consistently superior. The segmentation results produced by nnU-Net often included clearly visible errors that were not present in the segmentations produced with the proposed method.

Figure \ref{visualize_single} shows example segmentations for specific structures, which allow us to better visualize the segmentations errors and highlight under-segmentations and over-segmentations separately. Both our proposed method and nnU-Net segment most structures with good accuracy. Our method produces visibly superior segmentations on almost all structures and has less under-segmentation and less over-segmentation. Some of these structures, such as the cortical plate, have complex 3D geometries that are very difficult to segment manually. Although the additional errors in nnU-Net's output compared with our propsoed method may seem small, errors of this magnitude can make analyses which rely on consistent topology impossible and may necessitate long hours of manual correction. Our method produces better segmentations that can reduce the required manual corrections and enhance the accuracy and reproducibility of automatic analysis pipelines.

All methods considered for comparison above are based on fully convolutional networks. Further comparison with a non-DL method can be instructive and useful. To this end, we compared our proposed method with atlas-based segmentation, which is a popular classical method \cite{cabezas2011review, aljabar2009multi}. We used a multi-atlas segmentation (MAS) method similar to that described in Section \ref{data_and_annotation}, without the manual refinement, to segment the 22 test images. MAS approach segmented a target image via diffeomorphic deformable registration of at least 3 atlases to the target image, followed by label fusion using probabilistic STAPLE \cite{akhondi2013simultaneous}. The mean DSC of MAS segmentations on younger and older fetuses were, respectively, $0.875 \pm 0.102$ and $0.874 \pm 0.114$. These values were significantly ($p<0.001$) lower than those achieved by the proposed method presented in Table \ref{table:results}. The only structure that MAS segmented slightly more accurately than the proposed method was the cerebellum on the younger fetuses (mean DSC of 0.924 for MAS versus 0.921 for the proposed method that was statistically not significant, $p=0.30$). For all other structures, the proposed method achieved more accurate segmentation that were mostly statistically significant.

\begin{figure*}[!ht]
\centering
\includegraphics[width=0.88\textwidth]{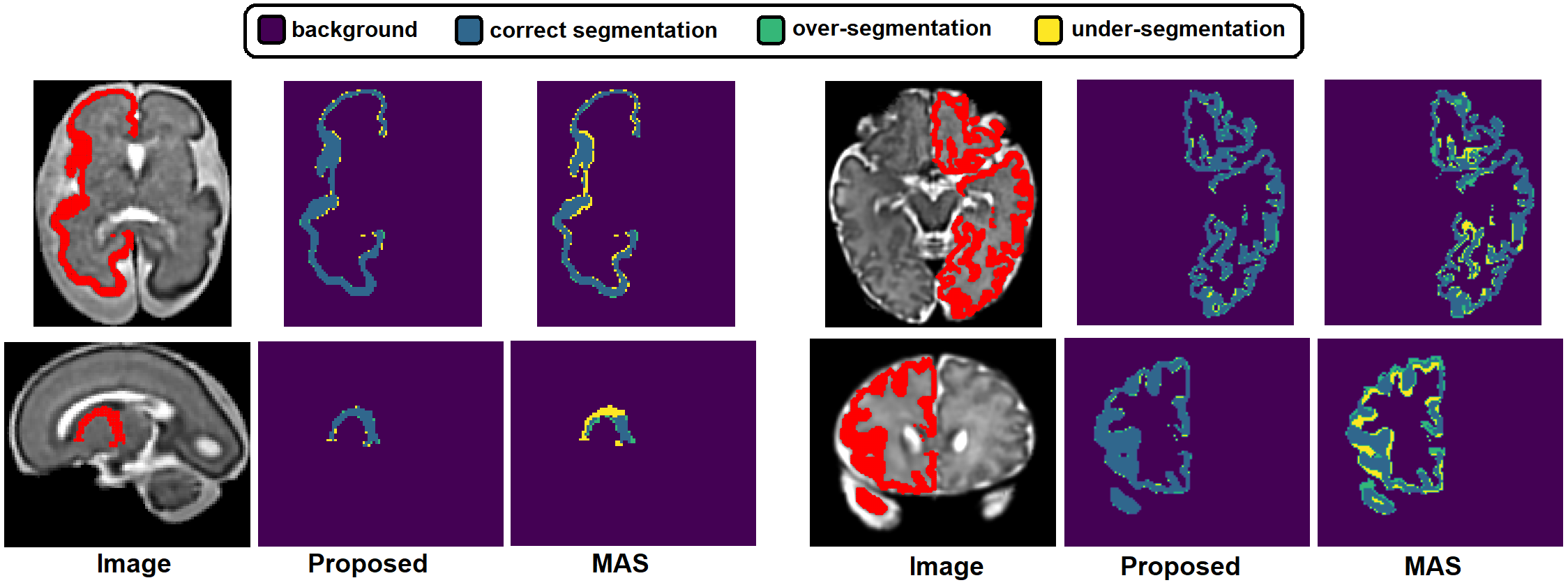}
\caption{Example segmentations predicted by our proposed method and multi-atlas segmentation (MAS). Each of the four examples in this figure shows one isolated structure and highlights correct segmentations, over-segmentations and under-segmentations.}
\label{visualize_MAS}
\end{figure*}

The difference between the proposed method and MAS were largest for more convoluted structures such as cortical plate (CP), subplate zone (SP), intermediate zone (IZ), ventricular zone (VZ), and white matter (WM). We show examples of the segmentations produced with the proposed method and MAS in Figure \ref{visualize_MAS}. For complex structures such as CP the segmentations produced by MAS show large errors. There are at least two causes for these errors. First, the registration between the atlas and the target images is never perfect, and the registration error is especially larger at the locations of thin and complex structures such as CP.  Registration is  difficult in fetal MRI due to stronger partial volume effects and reduced spatial resolution. Second, there is significant inter-subject variability in the local shape of these structures, and atlases cannot fully represent this variability. Therefore, regardless of the registration error, the segmentation accuracy of atlas-based methods is limited by the fact that an atlas can only represent the ``average shape'' and fails to capture the inter-subject variability. Deep learning methods avoid these pitfalls because they learn the complex relationship between the image intensity maps and the target segmentation map directly from the subject training data, rather than inferring it from an atlas.

The training time for our proposed method and all other deep learning methods was 30 hours. The only exception was training with the iMAE loss that required 100 hours. The inference time for a test image was approximately 4-6 seconds, depending on the brain size with larger brains taking closer to 6 seconds. For MAS, non-rigid registration of the atlas images to the target image took approximately 12 minutes and the label fusion with the probabilistic STAPLE took approximately 13 minutes, for a total of approximately 25 minutes.

\section{Conclusions}

As the diagnostic quality of fetal MRI improves and its applications grow, accurate quantitative analysis methods are increasingly needed to take full advantage of this imaging technique. Training DL models to address the needs of this application is challenged by the typically low image quality and by difficulty of obtaining large labeled datasets. Our study is a step forward in improving the accuracy and reproducibility of fetal MRI analysis as it enables accurate segmentation of the brain into more than 30 relevant and important tissue compartments. To address the main challenges outlined above, we used a semi-automatic method to obtain segmentation labels on our training images. This enabled us to segment a large number of training images with reasonable annotator time. We developed a novel training method based on a label smoothing strategy that accounted for tissue boundary uncertainties. This enabled us to account for the label noise in a systematic and principled manner in our model training. Our evaluations showed that our method produced significantly more accurate segmentations than the state of the art. The improved segmentation accuracy offered by our proposed method can translate into more accurate and more reproducible quantitative assessment of fetal brain development. It can also lead to substantial reductions in the annotator time because manual segmentations are very time-consuming. Finally, our methods may be useful in many similar setting in medical image segmentation where boundary uncertainties are important.

\section*{Acknowledgments}

This project was supported in part by the National Institute of Biomedical Imaging and Bioengineering, the National Institute of Neurological Disorders and Stroke, and the Eunice Kennedy Shriver National Institute of Child Health and Human Development of the National Institutes of Health (NIH) under award numbers R01EB031849, R01EB032366, R01HD109395, R01NS106030, K23NS101120, and R01NS121334; in part by the Office of the Director of the NIH under award number S10OD0250111; in part by the National Science Foundation (NSF) under award 2123061; in part by a research award from the Additional Ventures; in part by NVIDIA corporation through the Applied Research Accelerator Program and the Academic Hardware Grant Program; and in part by a Technological Innovations in Neuroscience Award from the McKnight Foundation. The content of this paper is solely the responsibility of the authors and does not necessarily represent the official views of the NIH, NSF, Additional Ventures, or the McKnight Foundation.

We used anatomical MRI scans from the DHCP in a transfer learning experiment. This dataset is provided by the developing Human Connectome Project, KCL-Imperial-Oxford Consortium funded by the European Research Council under the European Union Seventh Framework Programme (FP/2007-2013) / ERC Grant Agreement no. [319456]. We are grateful to the families who generously supported this trial.

\bibliographystyle{IEEEtran}
\bibliography{davoodreferences}

\end{document}